\documentclass{article}

\usepackage[preprint]{neurips_2020}

\usepackage[utf8]{inputenc} 
\usepackage[T1]{fontenc}    
\usepackage{hyperref}       
\usepackage{url}            
\usepackage{booktabs}       
\usepackage{amsfonts}       
\usepackage{nicefrac}       
\usepackage{microtype}      
\usepackage{graphicx}
\title{Application of variational policy gradient to atomic-scale materials synthesis}

\author{%
  Siyan Liu \\
  University of Kansas\\
  \texttt{} \\
   \And
   Nikolay Borodinov \\
   Siemens Corporation\\
   \texttt{} \\
   \AND
   Lukas Vlcek \\
   Joint Institute for Computational Sciences \\
   University of Tennessee, Knoxville \\
   \texttt{} \\
   \And
   Dan Lu \\
   Computational Sciences and Engineering Division \\
   Oak Ridge National Laboratory \\
   \texttt{} \\
   \And
   Nouamane Laanait\\
   Anthem, Inc. \\
   \And
   Rama K. Vasudevan\thanks{Send correspondence to this author. Code available at github.com/ramav87/KMC-SVPG} \\
   Center for Nanophase Materials Sciences \\
   Oak Ridge National Laboratory \\
   \texttt{vasudevanrk@ornl.gov} \\
}

\begin{document}

Notice: This manuscript has been authored by UT-Battelle, LLC, under Contract No. DE-AC0500OR22725 with the U.S. Department of Energy. The United States Government retains and the publisher, by accepting the article for publication, acknowledges that the United States Government retains a non-exclusive, paid- up, irrevocable, world-wide license to publish or reproduce the published form of this manuscript, or allow others to do so, for the United States Government purposes. The Department of Energy will provide public access to these results of federally sponsored research in accordance with the DOE Public Access Plan (http://energy.gov/downloads/doe-public-access-plan).

\newpage

\maketitle

\begin{abstract}
  Atomic-scale materials synthesis via layer deposition techniques present a unique opportunity to control material structures and yield systems that display unique functional properties that cannot be stabilized using traditional bulk synthetic routes. However, the deposition process itself presents a large, multidimensional space that is traditionally optimized via intuition and trial and error, slowing down progress. Here, we present an application of deep reinforcement learning to a simulated materials synthesis problem, utilizing the Stein variational policy gradient (SVPG) approach to train multiple agents to optimize a stochastic policy to yield desired functional properties. Our contributions are (1) A fully open source simulation environment for layered materials synthesis problems, utilizing a kinetic Monte-Carlo engine and implemented in the OpenAI Gym framework, (2) Extension of the Stein variational policy gradient approach to deal with both image and tabular input, and (3) Developing a parallel (synchronous) implementation of SVPG using Horovod, distributing multiple agents across GPUs and individual simulation environments on CPUs. We demonstrate the utility of this approach in optimizing for a material surface characteristic, surface roughness, and explore the strategies used by the agents as compared with a traditional actor-critic (A2C) baseline. Further, we find that SVPG stabilizes the training process over traditional A2C. Such trained agents can be useful to a variety of atomic-scale deposition techniques, including pulsed laser deposition and molecular beam epitaxy, if the implementation challenges are addressed.
\end{abstract}

\section{Introduction}

Reinforcement learning (RL) in recent years has achieved  impressive results in an array of problems in continuous and discrete action spaces, including in games such as Chess, Go, Atari \citep{silver2018general, mnih2015human}, as well as in  robotics \citep{devin2017learning}, such as a recent demonstration of using RL to solve a Rubik's cube puzzle \citep{mcaleer2018solving}. However, despite these considerable successes, applications outside of these 'traditional' domains remain limited, due to prohibitive sample inefficiency as well as a lack of available simulated environments on which agents can be trained for deployment.  This is particularly true in the domain of the physical sciences, where, although very good simulations exist for predicting static and dynamic systems ranging from simple molecules to complex proteins, solid-state matter and polymers, RL has made few inroads. Some notable exceptions include the use of deep RL in the case of molecular design and optimizing chemical reactions \citep{neil2018exploring}, and a recent report on use of RL for automating tip conditioning in scanning tunneling microscopy\citep{krull2020artificial}. As such, there is substantial potential to apply RL in such domains.

In this paper, we show the first application of RL to the case of atomic-level materials synthesis in a simulated environment. We developed a fully python-based kinetic Monte-Carlo model incorporating both atomic deposition and diffusion elements, and incorporated the environment into the  OpenAI Gym framework. Given that the simulations are necessarily expensive and highly stochastic, we utilized a recently developed variational policy gradient approach - Stein variational policy gradient - to train agents to optimize specific materials descriptors. We extended the existing algorithm by incorporating mixed image and tabular data during the training process, and developed a synchronous parallel implementation via Horovod that can be scaled to thousands of agents, with training occurring on GPUs. We then discuss the performance of the algorithm on the environment, and draw conclusions relevant for domain experts on the strategies employed by the agents. The paper is organized as follows. We begin with an overview of the specific problem from the domain side, and include a description of the underlying environment, as well as state and action spaces. Next, we introduce the Stein variational method as presented by Liu et al. \citep{liu2017stein} as well as a description of the modifications and parameters of the models. We then introduce the results, comparing the SVPG approach to a traditional actor-critic algorithm, and explore the robustness of the learned policies. Finally, we conclude with a discussion on the outlook of RL for materials synthesis, and core physical sciences more generally in light of these findings. 

\section{Simulation Environment}
\subsection{Kinetic Monte-Carlo}
The challenge we explore here is one of materials synthesis, specifically those of thin-films using atomic layer deposition approaches such as molecular beam epitaxy or pulsed laser deposition. These approaches have been pivotal in the past three decades in advancing our understanding of materials' structure and function, given that thin films can be engineered to possess a variety of properties unavailable through traditional bulk routes, and can be generated essentially free of extended defects and with exquisite control over levels of strain, defect density, and so forth \citep{christen2008recent}. Despite the proliferation of these deposition systems, control over the deposition process during the deposition itself remains limited for the most part. This is because of a limited ability to interpret the available signals that are available (typically, surface diffraction images) as well as the stochastic nature of film growth. As such, any interventions during a film deposition are typically confined to stopping flux of incoming atomic species for some time (to enable annealing)\citep{koster1999imposed}, or more basic operations such as changing the targets to enable growth of films with different compositional layers. As such, current methods are limited to quasi-static deposition conditions, which are not varied through the deposition process, and are time and labor intensive. Enabling automated synthesis with trained artificial agents would enable not only accelerated materials discovery, but also to potentially new states of matter that could be stabilized through unique policies that would be difficult if not impossible to discover through human trial and error. 

To explore this, we first created a simulated environment in which agents could be trained to optimize for a particular materials descriptor. The simulation utilizes a kinetic Monte-Carlo (kMC) engine and is loosely based on the simulation described in Tan et al. \citep{tan2005pulsed} The kMC simulation is a discrete lattice-based simulation that takes an input (starting) atomic configuration as well as rate parameters for distinct events that can occur, and then proceeds to sample from the events based on their probabilities, incrementing the simulation time in the process. More specifically, we outline five distinct events that can occur: (1) Deposition of atomic species of type $A$, (2) Deposition of atomic species of type $B$, (3) Diffusion of an atom into a neighborhood of similar atom types, (4) Diffusion of an atom into a neighborhood of different atom types, and (5) Diffusion of an atom into a neighborhood with mixed atom types.\footnote{In this case, 'same' refers to the neighborhood (nearest neighbors) being more than $\frac {3}{4}$ of the same type, different refers to nearest neighbors being less than $\frac {1}{4}$ the same type, while mixed refers to situations in between} We consider that atoms can be deposited on vacant sites on the surface, and further, that diffusion of atomic species also requires vacancies. A list is constructed of possible events that can occur, and the number of possible events of that type multiplied by the rate constant. A random number is chosen between (0,1) which determines the type of event from the constructed list, and a random atom (or atom site) is chosen to undergo the event chosen. The simulation clock is incremented as
\begin{equation}
    \Delta t = [\sum_{i=1}^{n} v_i ]^{-1} (-ln (R))
\end{equation}
where $v_i$ is the rate of the event $i$ and $ln (R)$, where $R$ is a random number on the interval (0,1) is added for mathematical completeness. There are two main points from this treatment: the first is that kMC, due to the parameters being rates, is a 'real-time' predictive simulation as opposed to some 'time step' type simulation, i.e. the actual times output by the simulation should be comparable to experiment. The second is that the probability of an event occurring is down to not only the rate constant itself, but also to the number of possible events of that type that can actually occur. For example, if there are no vacant sites for diffusion, then regardless of the input diffusion rate, no diffusion will be possible. Conversely, events with low rates, but with large numbers of possible sites which can undergo them, can be extremely frequent. 

\begin{figure}
    \centering
    \includegraphics[width=13.5cm]{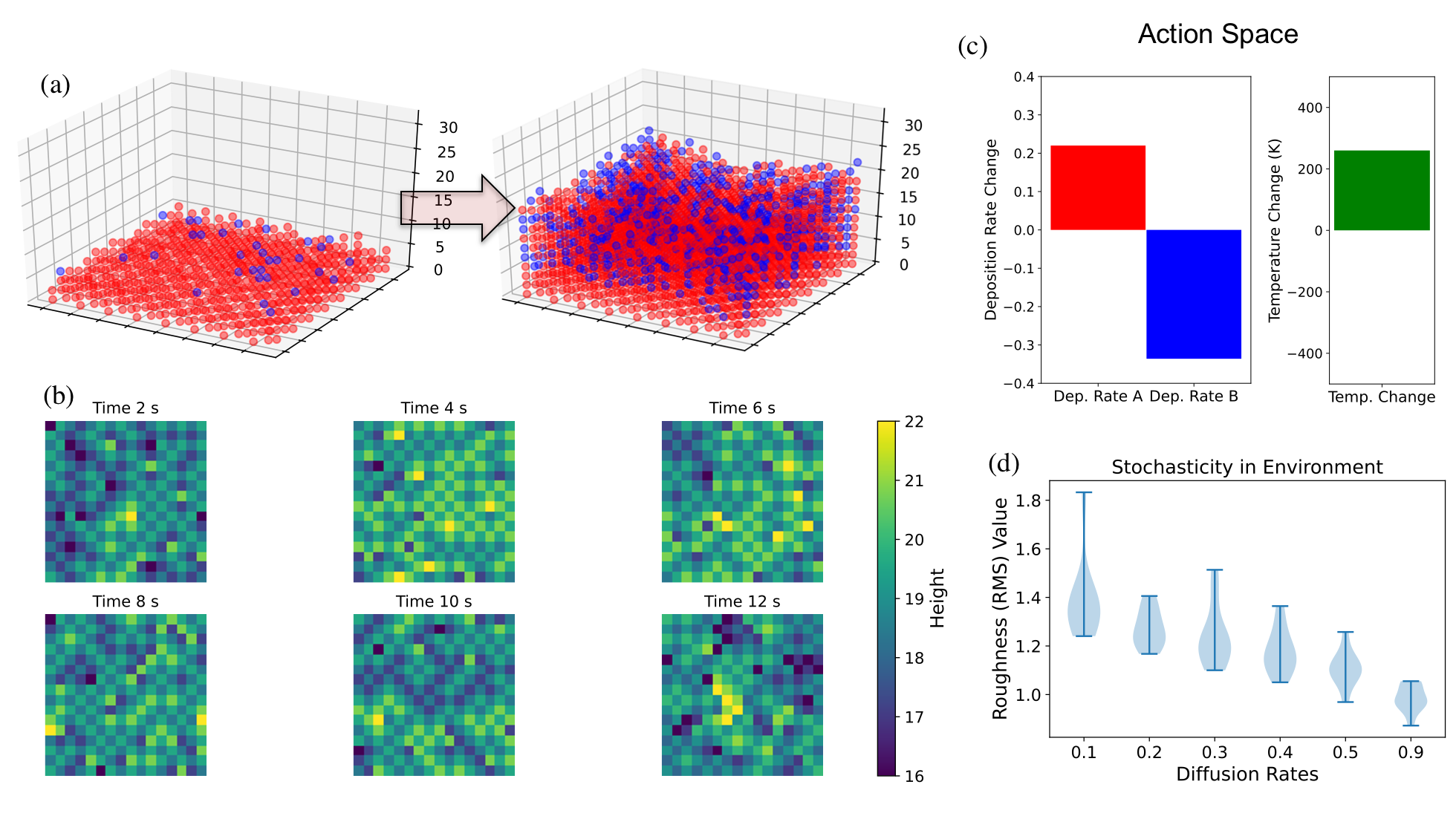}
    \caption{Simulation environment for materials synthesis. (a) Example of film growth progression, with atoms of two elements being colored in red and blue. (b) Surface projections at different time steps from the simulation in (a). (c) Action space for agents: increase and decrease deposition rates, and increase or decrease temperature. (d) Film growth simulation results for fixed deposition rates, and diffusion rates are given by the labels on the x-axis. The results show a distribution of roughness values corresponding to ten simulations run under identical conditions.}
    \label{fig:1}
\end{figure}

We implemented the simulation in pure python, assuming a face-centered cubic lattice crystal structure and a simulation box of size $[16,16,32]$ where the dimensions correspond to $(x,y,z)$ in real-space. An example of the output of the simulation is shown in Figure \ref{fig:1}(a), with the state of the film at 2 seconds and 10 seconds shown. Red and blue spheres correspond to distinct atomic species, called here for simplicity types $A$ and $B$, but these could be Ni, Co, Cu, etc. Note that the simulation begins with a single layer of atoms of a single type as can be seen in red, and subsequently the film growth process is initiated, enabling both atomic deposition and diffusion events to occur. This results in film growth and roughening of the surface. 

All of the rates for the five event types can be controlled independently in principle. In practice, deposition rates can be controlled independently in experiments, but diffusion rates are controlled through the choice of the material system and the conditions, typically the main factors being partial pressure of gases in the growth chamber, the kinetic energy of arriving species and the substrate temperature. For ease we consider here only the temperature, which will affect all three diffusion rates. We assume that each diffusion rate is governed by a linear relationship with the temperature, i.e. $D_i = m_iT + c_i$ where the values are given in Table ref{table:1} for the coefficients. At high temperatures, the diffusion rate is higher for atoms to move to different neighborhoods, whereas at lower temperatures the diffusion rates for diffusion into similar environments are highest, and the diffusion to different environments is the lowest. At ~560K there is a crossover where all three diffusion curves meet, i.e. the rates are the same regardless of the environment the atom is hopping into. 

Table 1: Diffusion rate dependence on temperature

\begin{center}
 \begin{tabular}{c c c} 
 \hline
 $i$ & $m_i$ & $c_i$ \\ [0.5ex] 
 \hline
 0 (Same)  & $4.19E-4$ & 0.215 \\ 
 \hline
 1 (Different) & $7.5E-4$ & 0.0280  \\
 \hline
 2 (Mixed) & $5.7E-4$ & 0.126  \\
 \hline
\end{tabular}
\label{table:1}
\end{center}

\subsection{State, Action Space and Rewards}
In general, a full 3D picture of the growth process will not be available in any real experiment - rather, the information will be limited to a surface projection if intermittent microscopy is performed ,\citep{voigtlander1993simultaneous} but dynamic information available during the growth process is limited to surface diffraction, typically in reflection geometry. Whilst this can include some information on sub-surface layers, the inversion of surface diffraction from reflection high-energy electron diffraction imaging is highly non-trivial, and even the forward simulations for arbitrary surface structures remains a vexing computational problem \citep{peng1996approximate}, that will not be explored here. Rather, we assume that the surface projection (2D image) is the state available to the agent, in addition to tabular data on the deposition rates, temperature, and surface fraction of atomic species of type $B$ on the surface. The latter can be derived from e.g., x-ray photoelectron spectroscopy measurements, or from Auger electron spectroscopy, or other forms of imaging. Thus, the state variable is \begin{math} \mathcal{S} = [S_{s}, M_{t}] \end{math} where $S_{s}$ is the 2D image of size $(16,16)$ and $M_t$ is a vector of values of length 4. An example of the progression of states during the simulation in Figure \ref{fig:1}(a) is seen in Figure \ref{fig:1}(b). 

The action space \begin{math} \mathcal{A} \end{math} (Figure \ref{fig:1}(c)) of the simulation   is continuous and consists of three distinct actions corresponding to the controls available: (1) The rate of deposition of atomic species $A$, (2) Rate of deposition of atomic species $B$, and (3) The temperature $T$. For the case of deposition, the rate can be changed by up to $\pm{0.25}$ for every intervention step, whilst the temperature can be altered by up to $\pm{500 K}$. The deposition rates are clipped at $(0.01, 0.90)$ and the temperature is limited to the window of $(300, 1400) K$. The simulation is initialized at $t = 0 s$, and then run until t $\approx$ 2 s, after which the deposition rates and temperature can be changed. The simulation runs for a further 2 s providing another intervention step, and so forth until the end of the simulation at $t\approx12 s$. Thus, six sequential actions occur per episode of the simulation. In terms of computational time, one complete simulation episode takes on the order of ~20 s on a typical CPU. This varies substantially, though, due to the length of the simulation depends on the particular simulation parameters chosen, and thus could be anywhere from ~ 10 - 40 s per episode.

A variety of reward functions can be considered, comporting to specific materials descriptors that are desired for targeted functional properties. These include specific structural features such as the presence of 3D mounding, step-type growth, surface roughness, and so forth, as well as chemical features such as level of surface segregation of specific atomic species, tendency towards formation of atomic clusters, etc. Here, we consider only the surface roughness as the material descriptor, but replacing it with other more complex descriptors is straightforward. The surface roughness $R_{film}$ is  measured as a root-mean-squared (RMS) value, effectively the difference between the individual height at an (x,y) position and the average height, normalized to the size of the surface image. We consider a reward function $R(\cdot)$ that takes as input the surface roughness, a target roughness that is desired, and then outputs a scalar value. In this , we consider a simple Gaussian centered around the target roughness, given at every intervention step in the simulation, but multiplied by 5 for the final state to force additional importance on the final state of the film, so $R(R_{film}, target)$ is

\begin{math}
   \left\{
\begin{array}{ll}
      
50e^{{{ - \left( {R_{film} - target } \right)^2 } \mathord{\left/ {\vphantom {{ - \left( {x - \mu } \right)^2 } {2\sigma ^2 }}} \right. \kern-\nulldelimiterspace} {2\sigma ^2 }}} -1
 & t < t_{f} \\
  250e^{{{ - \left( {R_{film} - target } \right)^2 } \mathord{\left/ {\vphantom {{ - \left( {x - \mu } \right)^2 } {2\sigma ^2 }}} \right. \kern-\nulldelimiterspace} {2\sigma ^2 }}} -5 & t = t_{f} \\
\end{array} 
\right. 
\end{math}

where the $target$ is the target roughness, and we set $\sigma = 0.0245$ and the simulation runs from $t = 0$ to $t = t_f$. This reward function has the effect of providing a -1 reward for any roughness substantially distant from the target roughness (-5 for the last step), on the order of 10\% or greater from the target value. 

The simulation is highly stochastic. To illustrate this, we show in Figure \ref{fig:1}(d) the roughness value of film growth simulated under the identical deposition rates but with the three diffusion rates set to equal a single value, ranging from 0.1 to 0.9, and plot the resulting distribution of roughness values for 10 individual simulation runs at each diffusion rate. This stochastic nature presents an inherent challenge for any agent in understanding and perturbing the system dynamics to realize desired material states. With the above state, action spaces and reward signals, the problem of material synthesis optimization can now be formulated in terms of standard supervised reinforcement learning terminology. 

\section {Reinforcement Learning}

Reinforcement learning offers a framework in which the aforementioned problem can be tackled. Briefly, RL concerns learning a policy $\pi$ to maximize cumulative rewards in a dynamic environment through repeated interactions. This can be expressed in terms of an optimization which seeks to maximize a utility function, which is the expected return of a policy $\pi$:
\begin{equation}
    J(\pi) = \mathbf{E}[\sum_{t=0}^{\infty}\gamma^t r(s_t, a_t)]
\end{equation}
where $\gamma$ is a discount factor, and actions $a_t$ are drawn from the policy $\pi(a_t|s_t)$, and the states $s_{t+1}$ are drawn from the dynamic environment, conditional on the action $a_t$ and the previous state $s_t$. We will assume here that the environment dynamics are unknown (model-free RL). By the policy gradient theorem, we may approximate the gradient of the utility function with respect to parameters $\theta$ that parameterize the policy via
\begin{equation}
    \nabla_{\theta} J(\pi) \approx \sum_{t=0}^{\infty}\nabla log \pi (a_t | s_t;\theta)R_t
\end{equation}
where $R_t = \sum_{i=0}^{\infty} \gamma^t r(s_{t+i}, a_{t+i})$ is the cumulative return from time $t$, and comprises the REINFORCE algorithm \citep{williams1992simple}. However, this method suffers from high variance, and convergence can be accelerated by subtracting a suitable baseline. The advantage actor-critic (A2C) algorithm \citep{sutton1998introduction} utilizes the value function baseline, 
\begin{equation}
    \nabla_{\theta} J(\pi) \approx \sum_{t=0}^{\infty}\nabla log \pi (a_t | s_t;\theta)(R_t - V^\pi (s_t))
\end{equation}
where the value function $V^\pi (s_t) = \mathbf{E}[\sum_{i=0}^{\infty} \gamma^i r(s_{t+i}, a_{t+i})]$ and gives the expected return for the agent from the state $t$, under the current policy $\pi$. Typically, both the policy and the baseline (value function or action-value function) are approximated using neural networks. 

\subsection{Stein Variational Policy Gradient}
We seek a reinforcement learning algorithm that will be both robust to different initialization, as well as sample efficient. Since each episode takes between 10 and 40 seconds, parallel methods to obtain information about the environment is required. Recently, Liu et al. \citep{liu2017stein} proposed a variational inference method to termed the Stein variational policy gradient (SVPG), where a set of policy particles ${\{\theta_i\}}$ is perturbed to achieve a balance between exploration (repulsion between policies) and exploitation (ascending the utility function). Specifically, the update rule they derived is
\begin{equation}
\Delta \theta_i \leftarrow \frac{1}{n} \sum_{j=1}^{n} [\nabla _{\theta _j} (\frac{1}{\alpha} J(\theta _j) + log q_0 (\theta _j)) k(\theta_j, \theta_i) + \nabla_{\theta _j} k(\theta _j, \theta _i)]
\end{equation}

It was further shown that, for simple continuous control problems, SVPG outperformed A2C in terms of time to solution (sample efficiency), and provided more state exploration than independent A2C actors. Given that this method satisfies the necessities of sample efficiency and (empirically observed) robustness, we chose this particular algorithm for our purpose.

\section {Results}
\begin{figure}
    \centering
    \includegraphics[width=13.5cm]{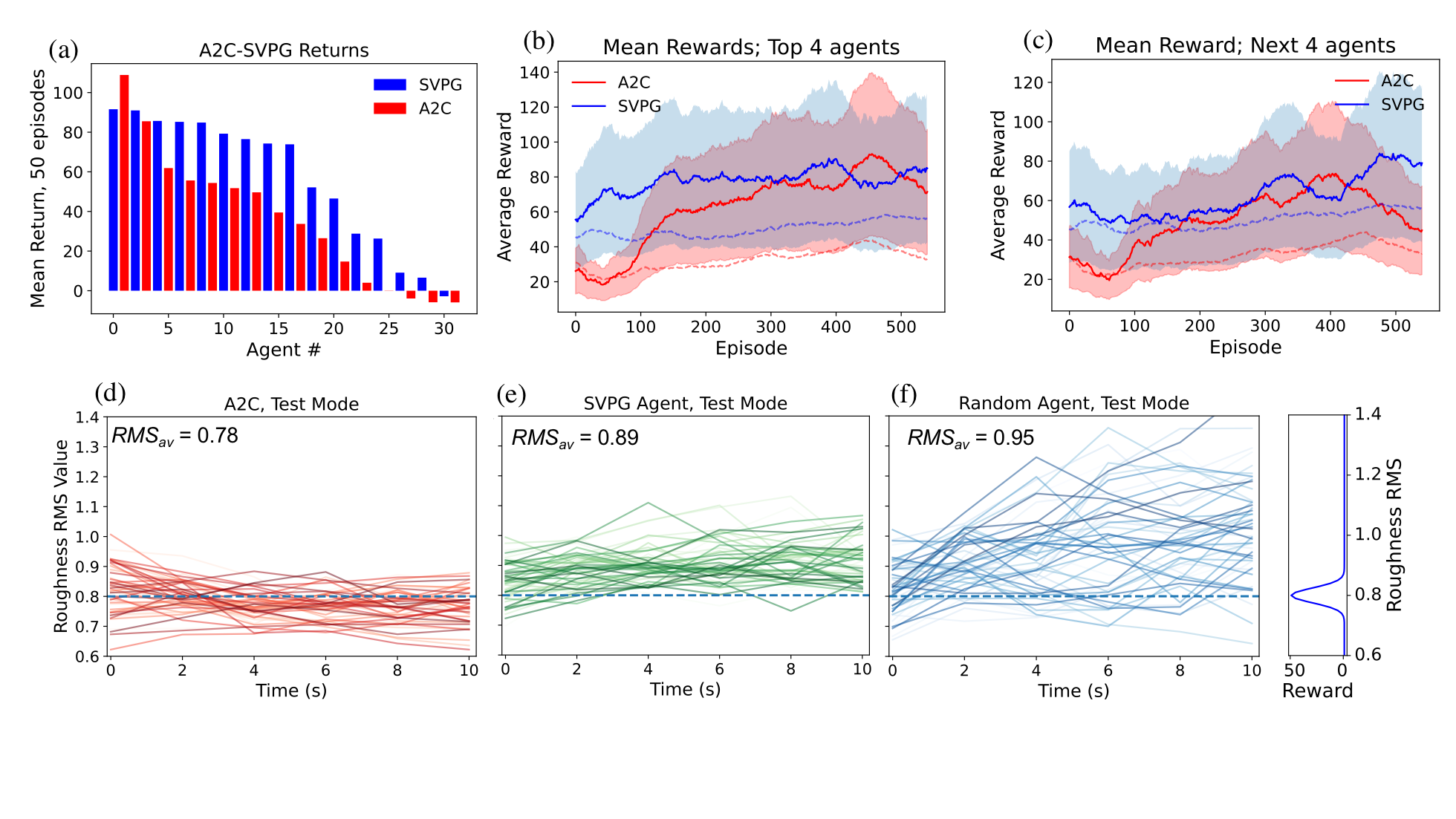}
    \caption{(a) Results, with average returns from the last 50 episodes of training for the SVPG and A2C agents. Mean rewards for the top four agents (b) and the next four agents (c) for both methods as a function of training. Note that smoothing has been used with a convolutional filter of size 60, and shaded regions are half a standard deviation. (d-f) Test results from 50 runs, of the best-performing agents, compared with results from a random agent. The reward function is shown on the right.} 
    \label{fig:2}
\end{figure}

\begin{figure}
    \centering
    \includegraphics[width=13.5cm]{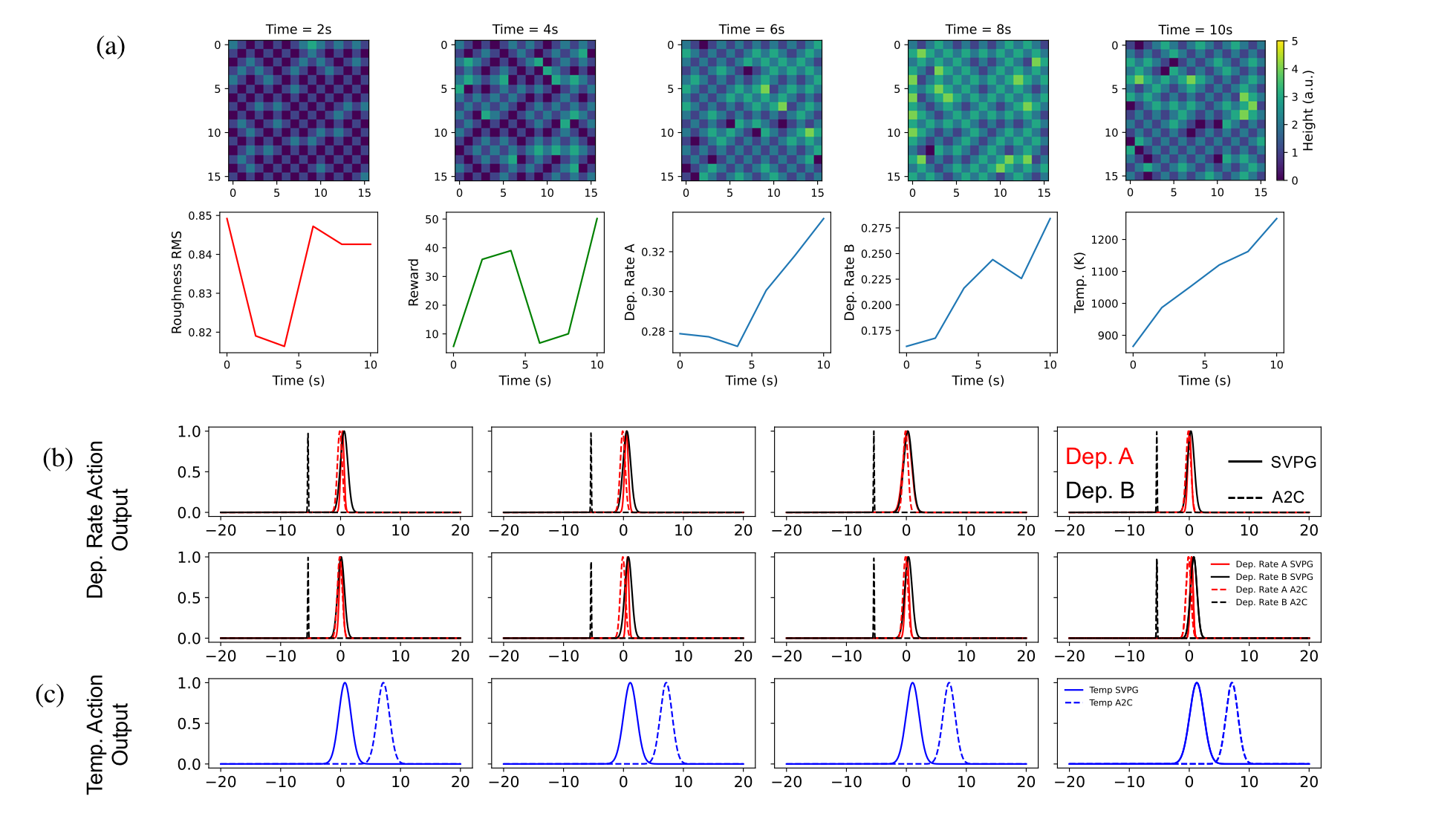}
    \caption{(a) An example of a simulation run with the surface projections of the film at different time steps shown above, and the roughness, rewards, deposition rates and temperatures below made by the SVPG agent. (b) Policy outputs for randomly selected states passed through the actor networks for both A2C (dashed lines) and SVPG (solid line) agents. Plotted here are normal distributions with parameters $\mu$ and $\sigma$ given by the respective networks. Overlap of the distributions suggests that the same action is preferred by both agents; larger variance indicates somewhat more exploration by the agent. The output for temperature actions is shown in (c) in the same manner. } 
    \label{fig:3}
\end{figure}

In order to apply SVPG for our method, we first had to extend the existing SVPG implementation to allow for more complex network architectures. Our architecture is shown in supplementary materials and consists of two separate inputs corresponding to the distinct data types present: a 2D image is fed through convolutional layers, while the vector of tabular data on rates, temperature and atomic fraction are fed into fully connected layers. These two branches are concatenated and then fed through two more fully connected layers. Since the model is an actor-critic model, the agent consists of both an actor model and a separate critic model. The structure of both are identical in our setup, with the exception of the output layer. The full network architecture is included in the supplemental material. For the actor model the output layer is size (6), corresponding to predictions of the mean $\mu$ and standard deviation $\sigma$ of a diagonal Gaussian policy from which the three actions are sampled. Linear activations are used on the last layer for both models. For the critic model, the output is a single scalar value, corresponding to the estimate of the value function for that state. Note that for the actor model, the standard deviation is ensured to be positive by passing through a softplus layer on the output, i.e. if the output of the standard deviation is $\sigma'$, then $\sigma=log(exp(\sigma')+1)$. Actions sampled from the diagonal Gaussian policy are clipped to lie in the interval [-10,10] and linearly scaled to reflect changes to deposition rates (that lie between [-.25, .25] and temperature [-500 K, +500 K]. The new deposition rates and diffusion rates are calculated based on the sampled action, clipped according to minimum and maximum allowable values, and then the simulation is run to the next intervention step. 

As in the original paper by Liu et al. \citep{liu2017stein}, we did not apply the Stein update to the critic, rather only to the actor models. We compared the policies trained with pure A2C, SVPG and benchmarked against a random agent. Finally, we developed a synchronous parallel version of SVPG that could scale to multiple nodes on high performance computing systems, via the Horovod package \citep{sergeev2018horovod}. We situated multiple agents on GPUs and all simulations on CPU processes. After every 20 episodes, an \emph{allgather} is performed to obtain the policy gradient loss and actor network weights from each agent. The Stein update is then calculated based on the gathered gradients and weights, and then the individual policies are updated in the direction of the Stein gradient. We trained for 400 episodes in total for both the standard A2C and SVPG methods. We utilized the Adam optimizer for the Stein updates with a learning rate of $10^{-2}$, and the temperature parameter was $\alpha$ = 5.0. For the both methods, the Adam optimizer was used for both actor and critic updates with a learning rate of 0.005 for the actor and 0.01 for the critic. Finally, the discount factor used in all cases was $\gamma = 0.90$. All training was done on a single nVIDIA DGX server with 4 Tesla V-100 GPUs. A single run of training took approximately 2 hours. Note that this can be reduced by an order of magnitude if we parallelize the simulations run for each batch. This work is ongoing.

We compare the returns of all agents trained with both methods in Figure \ref{fig:2}(a), sorted in descending order. Here, the return is taken as the average of the last 50 episodes during training. There are two important points to note: first, the best A2C agent outperforms the best SVPG agent. Secondly, the SVPG-trained agents show much less variance in the returns than the A2C method. Even though the average return is lower, they are uniformly higher than the corresponding A2C agent returns for the third-best agent onward. This is also well reflected in the training curves, seen in Figure \ref{fig:2}(b,c) for the top 4 agents and the next top four agents, respectively. The SVPG training curve in both plots increases steadily whereas the opposite is true for the A2C agents in Figure \ref{fig:2}(c). The mean of all agent returns is shown as dashed lines in both plots. Note that smoothing has also been applied with a convolutional filter of size 60. 

Next, we aimed to test the agents and plot the corresponding film roughness as a function of time, for 50 different simulations. These results are shown in Figure \ref{fig:2}(d-f), with the reward function seen on the right. Each agent aims to keep the film roughness as close as possible to the dashed blue line drawn. The best-performing A2C agent was compared with the best-performing SVPG agent, and finally, with the performance of a random actor. The best-performing A2C actor effectively squashes the distribution to be tight around the target roughness, although more-so on the lower side, even though the reward function is symmetric. On the other hand, the SVPG agent appears to be slightly more varied, with a larger density on the upper  side, as well as several 'runaways' where the roughness increased well away from the target. This still compares favorably to the actions of the random agent. The mean roughness values for the A2C agent was 0.78 over all steps for all runs, compared with 0.89 for the SVPG agent, and finally 0.95 for the random agent. These results clearly highlight that the agents, whether trained through SVPG or A2C, are developing strategies to control the material characteristic specified. 

\subsection{Policy Inspection}
Of importance are the specific strategies that are being used by the agents to reduce the surface roughness. In general, film growth will lead to increased roughness over time due to the random nature of the surface deposition and diffusion events combined with limited time for diffusion in the presence of energy barriers (e.g., consider mounding instabilities \citep{pierre1999edge, stroscio1995coarsening, lengel1999nonuniversality}. One option is to enable the surface to reconstruct after deposition for some time, in a technique called pulsed laser interval deposition \citep{koster1999imposed}. Although this particular action is not available to the agent, it would be equivalent to dramatically lowering the deposition rates and increasing the temperature to encourage more surface diffusion. Observing an example in Figure \ref{fig:3}(a), the SVPG agent does indeed reduce the deposition rate for atom species A to the minimum, but surprisingly maintains or increases slightly the deposition rate throughout the process. Meanwhile, the temperature is kept to steadily increase. This is a somewhat unusual strategy from the point of view of domain expertise, but apparently led to a large reward. At higher temperatures, the diffusion of atoms into mixed neighborhoods is enhanced compared to diffusion to similar atom neighborhoods; thus it could be beneficial to increase the deposition rate slightly to 'correct' for clusters forming that can increase roughness. 

Since the policies themselves are diagonal multivariate Gaussian policies, one method of inspecting the degree of exploration is to simply observe the mean and variance outputs for different states. Shown in \ref{fig:3}(b,c) are outputs for both SVPG and A2C agents for randomly selected states taken from during the test runs. Solid lines are for the action output of the best-performing SVPG agent, while the dashed lines are for the best-performing A2C agent. Interestingly, both of the policies appear to have similar actions for the first deposition rate, shown in red. However, while the SVPG agent has a slightly broader distribution of actions for the second deposition rate, the A2C agent has a sharp delta function to reduce the deposition rate in all the states considered. The variances for the temperature action are similar for both policies, but again the SVPG agent is more conservative than the A2C, likely reflecting the combined experience from multiple agents that temper the gradient update. 

\section{Discussion and Future Work}

The results presented in this work suggest that reinforcement learning agents could be utilized, in principle, to discover novel strategies for atomic-level material synthesis beyond the existing routines. The ability to utilize agents that are both robust, sample efficient, and adaptable to real world settings remains a challenge as reinforcement learning shifts to these new domains, where data is inherently expensive to acquire. Numerous challenges, however, must be addressed to make this a reality. First, the relatively simple environment simulated here must be extended to allow for enumerating many more types of events, and  their rates should be fit based on either first principle simulations, experiments, or a combination of the two, so that the environment can faithfully represent the true film deposition situation. This will present scalability problems as simulation times run longer and system sizes increase. Next, the state function will need to be modified via inversion of the surface diffraction images, or at least a rapid forward model for arbitrary surface structures which remains  difficult \citep{peng1996approximate}. Future work should therefore be to bring the environment closer to the realistic setting, as well as incorporate model-based reinforcement learning within the SVPG approach, such that the agents should be able to rapidly learn during individual synthesis runs. As reinforcement learning steadily develops, and lab automation in materials science gains prominence, these methods can be expected to lead to new methods of synthesizing matter.


\begin{ack}
\texttt{This research was funded by the AI Initiative, as part of the Laboratory Directed Research and Development Program of Oak Ridge National Laboratory, managed by UT-Battelle, LLC, for the U.S. Department of Energy (DOE). The core of the simulation was supported by the U.S. Department of Energy (DOE), Office of Science, Basic Energy Sciences (BES), Materials Sciences and Engineering Division (LV). A portion of this work was performed at the Oak Ridge National Laboratory’s Center for Nanophase Materials Sciences (CNMS), a U.S. DOE Office of Science User Facility.}
\end{ack}

\bibliographystyle{apa}
\bibliography{references.bib}

\end{document}